\documentclass[preprint,review,12pt,numbers,sort&compress]{elsarticle}

\usepackage{amssymb}
\usepackage[breaklinks=true]{hyperref}
\usepackage{breakurl}

\usepackage{epstopdf}

\journal{Physics Letters A}

\begin{document}
\begin{frontmatter}


\title{Collisional-Radiative Model for the visible spectrum of  W$^{26+}$ ions}


\author[nwnu]{Xiaobin Ding\corref{dd1}} \ead{dingxb@nwnu.edu.cn}
\author[nwnu]{Jiaxin Liu}
\author[SU]{Fumihiro Koike}
\author[NIFS]{Izumi Murakami}
\author[NIFS]{Daiji Kato}
\author[NIFS]{Hiroyuki A Sakaue}
\author[ECU]{Nobuyuki Nakamura}
\author[nwnu]{Chenzhong Dong}
\cortext[dd1]{Corresponding author.}

\address[nwnu]{Key Laboratory of Atomic and Molecular Physics and Functional Materials
of Gansu Province, College of Physics and Electronic Engineering,Northwest Normal University, Lanzhou 730070, China}

\address[SU]{Department of physics,Sophia university,Tokyo,102-8554,Japan}

\address[NIFS]{National Institute for Fusion Science, Toki, Gifu 509-5292, Japan}

\address[ECU]{Institute for Laser Science, The University of Electro-Comunications,Chofu,Tokyo 182-8585, Japan}

\begin{abstract}

Plasma diagnostics in magnetic confinement fusion plasmas by using visible spectrum strongly depends on the knowledge of fundamental atomic properties. A detailed collisional-radiative model of W$^{26+}$ ions has been constructed by considering radiative and electron excitation processes, in which the necessary atomic data had been calculated by relativistic configuration interaction method with the implementation of Flexible Atomic Code. The visible spectrum observed at an electron beam ion trap (EBIT) in Shanghai in the range of 332 nm to 392 nm was reproduced by present calculations. Some transition pairs of which the intensity ratio are sensitive to the electron density were selected as potential candidate of plasma diagnostics. Their electron density dependence are theoretically evaluated for the cases of EBIT plasmas and magnetic confinement fusion plasmas.

\end{abstract}

\begin{keyword}
 tungsten, visible spectrum, collisional-radiative modeling, plasma diagnosis,
\end{keyword}

\end{frontmatter}


\section{Introduction}
Tungsten(W) was chosen to be the cover material for the first-wall and divertor in the next generation magnetic confinement fusion (MCF) reactors, such as ITER, ASDEX, and EAST, due to its favourable physical and engineering properties such as low sputtering, high melting point, and low tritium retention rate\cite{0029-5515-45-3-007,Matthews2009934}. However, tungsten impurity ions might inevitably be produced during the interaction between the edge plasma and cover material. These ions may be transported to the fusion core plasma, and be ionized further to produce highly charged W ions. These highly charged W ions will undergo radiative decay by emitting high energy photons. Consequently, large radiation loss could be caused by these highly charged impurity ions, which will lead to plasma disruption if the relative concentration of W ions in the core plasma is higher than about 10$^{-5}$\cite{Radtke2007}. Monitoring and controlling the flux of these highly charged W impurity ions will be important to retain the fusion\cite{1402-4896-2009-T134-014022}. Thus a thorough knowledge of atomic properties of tungsten ions will be helpful for MCF research. Nevertheless, tiny amount of tungsten impurity ions can also provide us with plenty of information about fusion plasmas such as electron density, electron temperature and ion temperature by their spectra. Thus, highly charged W impurity ions could be used for diagnosis of fusion plasmas. For the diagnostics with spectra, it is necessary to investigate the fundamental properties W ions such as energy levels, radiative transition probabilities \emph{etc.}.

In the last three decades, extensive work had been performed on the atomic structure and radiative transition properties of highly charged W ions\cite{1402-4896-2011-T144-014011, 0953-4075-43-14-144013, PhysRevA.90.052517, 11, 0953-4075-43-20-205004}. Most of experimental works have been carried out by using electron beam ion traps (EBITs) and fusion reactor facilities. An EBIT is a widely used device which can selectively produce and trap highly charged ions with specific ionization stage by using a mono-energetic electron beam. Since the ions can be trapped in an EBIT for relatively long time and the electron density in EBIT is relatively low, it is possible to observe weak transitions from those highly charged ions such as magnetic dipole(M1) or electric quadrupole (E2) transitions. Komatsu et al.\cite{AkihiroKOMATSU2012} observed the visible spectrum of highly charged W$^{8+ - 28+}$ ions by using an compact electron beam ion trap, called CoBIT, in Tokyo\cite{Nakamura2008}. Several lines from M1 transitions among the ground state multiplets of corresponding ion were identified. The visible M1 transition of W$^{26+}$ ions was firstly observed by them\cite{1402-4896-2011-T144-014012}. Yanagibayashi \emph{et al.}\cite{0953-4075-43-14-144013} observed Extreme Ultra Violet (EUV) spectra in 2.6-3.2nm range of highly charged W ions from JT-60U plasma at $T_e \approx$ 8 and 14keV which are identified to be of the $3p\to3d$ transition from W$^{47+}$ to W$^{54+}$ ions. Z. Fei \emph{et al.}\cite{PhysRevA.90.052517} observed spectrum of the forbidden transition of W$^{26+}$ with a compact electron beam ion trap in Shanghai (SH-Perm EBIT). Radtke \emph{et al.}\cite{11} observed the emission spectra of W ions with charge states 21 to 50 for the wavelength ranges 4.5 - 7.0nm and 0.5 - 0.62nm with the Berlin EBIT. Harte \emph{et al.}\cite{0953-4075-43-20-205004} have studied the EUV spectra of tungsten with Large Helical Device(LHD) in National Institute for Fusion Science (NIFS), Japan. The spectra have been analyzed by using quasi-relativistic theory.

  For theoretical study of the level structure and radiative transition properties of the highly charged tungsten ions, the inclusion of relativistic effect and the electron correlation effects is indispensable since tungsten ions are of the heavy and multi-electron system. Various theoretical methods were used to calculate the properties of highly charged W ions\cite{PhysRevA.90.052517,PhysRevA.82.014502,PhysRevA.83.032509,PhysRevA.83.032517,0953-4075-44-14-145004}. Z. Fei \emph{et al.}\cite{PhysRevA.90.052517} calculated the M1 transitions among the ground-state configuration of W$^{26+}$ with multi reference relativistic many-body perturbation theory (MR-RMBPT). Gaigalas \emph{et al.}\cite{PhysRevA.82.014502,PhysRevA.83.032509} made a large-scale calculation using multi-configuration Dirac-Fock (MCDF) theory on the E1 and E3 transitions of W$^{24+}$ ions in both relativistic and non-relativistic limits taking the valence-valence and core-valence correlation effects into account. Ralchenko \emph{et al.} \cite{PhysRevA.83.032517} measured the M1 transitions of 3d$^{n}$ configurations in super EBIT, and the spectrum was analysed by detailed collisional-radiative (CR) modelling. Xiao-Bin Ding \emph{et al.}\cite{0953-4075-44-14-145004} calculated the M1 visible transitions among the ground state multiplets of the W$^{26+}$ ion using MCDF method. Both valence-valence and core-valence correlation effects had been taken into account in the calculation of the energy level and transition probability. One of the M1 transition lines between $^3H_5 \to  {^3H_4}$ has been assigned to the peak at 389.4nm which was observed by Komatsu \emph{et al.}\cite{AkihiroKOMATSU2012}. Furthermore, strong transition lines between $^3H_6 \to {^3H_5}$ and $^3F_3 \to {^3F_2}$ have been predicted theoretically, which have been observed by the experiment with CoBIT  at 464.4 nm and 501.9 nm, respectively\cite{0953-4075-44-14-145004}.

  The present paper is mainly focus on the emission spectrum and the intensity ratio of M1 transitions from W$^{26+}$ ground state multiplets which might be used as potential candidate of diagnosis lines. A detailed collisional-radiative model was constructed to investigate the spectra from highly charged ions in EBIT and fusion plasmas by assuming the electron energy distribution function as a $\delta$ function and Maxwellian, respectively.

  \section{Theoretical method}

  The CR model has been used successfully in many previous studies to analyze and identify dozens of EBIT spectral lines from highly charged heavy ions in X-ray and VUV as well as optical regions\cite{Ralchenko2013,0953-4075-48-14-144028}. In CR model the plasma was assumed to be optically thin and isotropic. The ground state of W$^{26+}$ is in  4d$^{10}4f^2$ configuration which have two electrons in the $4f$ subshell, and consequently have complex electron correlation effect. The atomic data, such as energy levels, radiative transition rates, and cross sections of collisional (de)excitation, which are necessary to construct the CR model, can be obtained in the framework of full relativistic configuration interaction (RCI) method with the implementation of Flexible Atomic Code (FAC) packages\cite{gu2008}. Based on these atomic data, a CR model code has been  developed to investigate the spectra from EBIT.

 The emission line intensity $I_{p,q}(\lambda)$ due to a radiative transition with wavelength $\lambda$ from the upper level $p$ to the lower level $q$ in an optically thin plasma, can be defined as:
\begin{equation}\label{eq1}
  I_{p,q}( \lambda) \propto n(p) A(p,q) \phi(\lambda), \label{eq1}
\end{equation}
where $n(p)$ is the population of the ions in the upper level \emph{p}, $A(p,q)$ is the transition probability or Einstein coefficient for transition from \emph{p} to \emph{q}, and $\phi(\lambda)$ is the normalized line profile. In this work, $\phi(\lambda)$ was taken as a Gaussian profile, which may include the effects of Doppler, natural, collisional and instrumental broadenings. The quantities $ A(p,q)$ can be obtained from the experiments or accurate theoretical calculations. The populations $n(p)$, on the other hand, are determined by various atomic process such as spontaneous radiative transition, (de)excitation by electron collisions, radiative recombination, dielectronic recombination, ionization by electron impact, or three-body recombination, while the effect of radiative and dielectronic as well as three body recombination processes is expected to be negligible in the cases of current interest, because they scarcely affect the population of the low-lying levels that are relevant to the EBIT spectrum\cite{0953-4075-45-3-035003}.

The temporal development of the population $n(p)$ in level \emph{p} is described by the following rate equation:
    \begin{eqnarray}
\frac{{\rm d}}{{\rm d} t}n(p)=&&\sum_{q<p}C(q,p)n_en(q) \nonumber \\
                              &&-[{\sum_{q<p}F(p,q)n_e+A(p,q)} + \sum_{q>p}{C(p,q)n_e}]n(p) \label{eq2}\\
                              &&+\sum_{q>p}[F(q,p)n_e+A(q,p)]n(q) \nonumber
\end{eqnarray}
 where ${n_e}$ is the electron density, $C(q,p)$ and $F(q,p)$ are collisional excitation and deexcitation rate coefficients from the level \emph{q} to \emph{p}, respectively. They can be calculated from the cross sections of collisional (de)excitation processes by assuming an appropriate free electron energy distribution. The electron energy distribution in the electron beam of an EBIT is mostly mono-energy under typical operation conditions\cite{PhysRevA.83.032517}. Thus, in the present work for the EBIT case, the electron energy distribution function is taken as a $\delta$ function, while the Maxwellian distribution of the electron energy was assumed in the case of fusion plasmas. The first term in the righthand side of equation (\ref{eq2}) refers to the population flux by the excitation processes from energy levels lower than \emph{p} and the last term also represents the population flux by collisional deexcitation and radiative transition processes from the levels higher than \emph{p}. The second term represents the depopulating flux of  level \emph{p} by both collisional excitation, deexcitation and radiative transition processes. A Quasi-Steady-State (QSS) equilibrium approximation was assumed, in which we set $ {\rm d}n(p)/{{\rm d} t}=0$\cite{1402-4896-2011-T144-014012}. Under this approximation, ${n(p)}$ can be solved from the set of equations (\ref{eq2}) to obtain the intensity $I_{q,p}( \lambda)$ by equation (\ref{eq1}).

 In order to construct the CR model for the W$^{26+}$ ions, the configurations $4d^{10} 4f^{2}$, $4d^{10} 4f^{1} nl$ ($n$ = 5,6,7,8), $4d^{10} nl n'l' $ ($n,n'$=5,6), $4d^{9} 4f^{2} nl$ ($n$=5,6,7) and $4d^{9} 4f^{1} nl n'l'$ ($n,n'$=5,6) with \emph{l} and \emph{l'} from 0 to \emph{n}-1 and \emph{n'}-1 correspondingly, were included in the present calculation. All the energy levels, radiative transition probabilities and collision excitation cross sections among these levels were calculated. Most of the important configuration interaction effects were included in the present calculation.

\section{Results and discussion}

  The level energies of the ground state $4d^{10}4f^{2}$ multiplets of W$^{26+}$ have been calculated and are given in Table \ref{tab1}. The first column is the state designation in the LS coupling scheme. All the columns labeled as Cal(nl) are the level energies calculated from different electron correlation models. All the important configuration interaction contribution was included in a systematical way by constructing the interacting configurations from the single and double subsititution from \emph{n}=4 subshells of W$^{26+}$ ions to \emph{nl} virtual orbits, where \emph{nl} means the highest principle quantum number \emph{n}(\emph{n}=5,6,7,8) and angular quantum number \emph{l}(\emph{l}=0,1,\dots,(n-1)) of excited electron. The available calculated result by multi reference relativistic many-body perturbation theory(MR-RMBPT) and experimental observation by EBIT\cite{PhysRevA.90.052517} were also provided. It can be inferred from the table that the level energies get converged by the increase of $nl$. The result of the present calculation agrees well with other calculations and experiments.

\begin{table}
\begin{center}
\caption{Energy levels in eV of the 4f$^2$ configuration of W$^{26+}$ ions from relativistic configuration interaction   calculations with the FAC code. The column of 'Cal(nl)' refer to different configuration interaction effects included in the calculation. The interacting configuration was constructed by the single and double subsititution from \emph{n}=4 subshells of W$^{26+}$ ions to \emph{nl} virtual orbits where \emph{nl} means the highest principle quantum number \emph{n}(\emph{n}=5,6,7,8) and angular quantum number \emph{l}(\emph{l}=0,1,\dots,(n-1)). The energy of the ground state level was set to zero. The column 'Ref' and 'Exp' gives the theoretical result of multi reference relativistic many-body perturbation theory(MR-RMBPT) and experimental observation by Z. Fei et al\cite{PhysRevA.90.052517}.
}\label{tab1}
\vspace{2mm}
\begin{tabular}{p{1.5cm}p{1.5cm}p{1.5cm}p{1.5cm}p{1.5cm}p{1.5cm}p{1.5cm}}

\hline
{LS}       & {Cal(5l)}    & {Cal(6l)}  & {Cal(7l)}   & {Cal(8l)}        & {Ref}       & {Exp}  \\\hline
{$^3H_4$}  & {0}            & {0}        & {0}         & {0}            & {0}          & {0}      \\
{$^3F_2$}  & {2.3054}       & {2.3555}   & {2.3242}    & {2.2192}       & {2.2393}     & {2.2182}  \\
{$^3H_5$}  & {3.2324}      & {3.2111}   & {3.2126}    & {3.1815}       & {3.1891}     & {3.1836}  \\
{$^3F_3$}  & {4.7917}       & {4.8148}   & {4.7668}    & {4.6897}       & {4.6933}     & {4.6874}  \\
{$^1G_4$}  & {4.7257}      & {4.7473}   & {4.7388}    & {4.7117}       & {4.6972}     & {4.7095}  \\
{$^3H_6$}  & {5.9540}       & {5.9312}   & {5.9322}    & {5.8541}       & {5.8540}     & {5.8521}  \\
{$^3F_4$}  & {8.4873}      & {8.4965}   & {8.4875}    & {8.4029}       & {8.4073}     & {8.4022}  \\
{$^1D_2$}  & {8.7858}      & {8.9140}   & {8.8840}    & {8.4845}       & {8.4846}     & {      }  \\
{$^3P_0$}  & {9.0766}      & {9.2754}   & {9.2337}    & {8.7099}       & {8.7335}     & {      }  \\
{$^3P_1$}  & {10.6522}     & {10.8267}  & {10.7923}   & {10.2943}      & {10.2852}    & {      }  \\
{$^1I_6$}  & {11.2278}     & {11.3674}  & {11.3594}   & {10.7850}      & {10.7883}    & {      }  \\
{$^3P_2$}  & {13.0979}      & {13.2399}  & {13.2065}   & {12.7467}      & {12.7229}    & {      }  \\
{$^1S_0$}  & {21.8196}      & {22.1866}  & {22.0620}   & {21.5404}      & {21.5223}    & {      }  \\
\hline
\end{tabular}
\end{center}
\end{table}

\begin{center}
\begin{figure}
\includegraphics {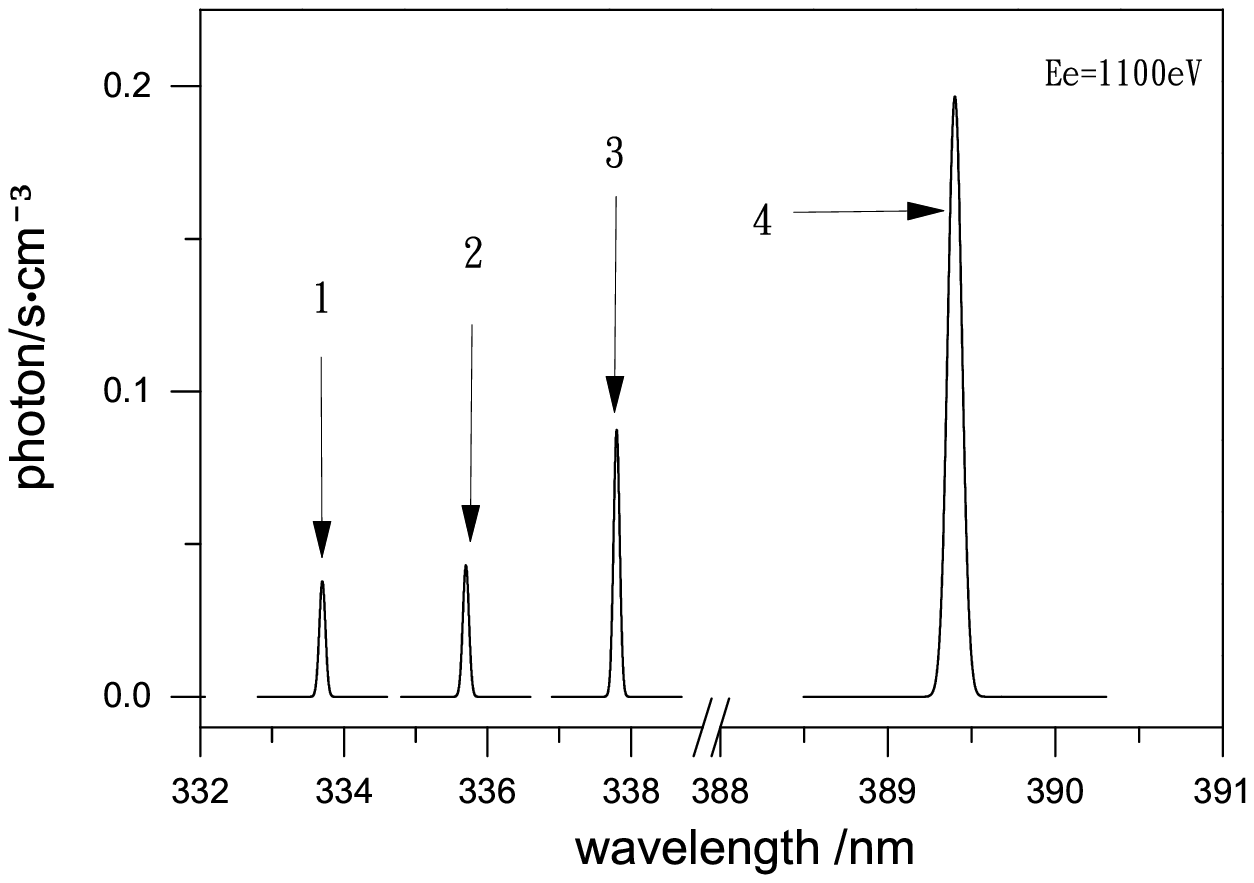}
\caption{The synthetic visible M1 spectrum of W$^{26+}$ ions at the electron beam energy of 1100eV with EBIT in 332-340nm and 388-391nm. The peaks marked as 1, 2 and 4 are the M1 transitions $^3F_4\to{^3F_3}$, $^3F_4\to{^1G_4}$, and $^3H_5\to{^3H_4}$ from W$^{26+}$ ions, respectively. The peak 3 is the M1 transition $^2F_{5/2} \to {^2F_{7/2}}$ of W$^{27+}$ ions. The population ratio between W$^{26+}$ and W$^{27+}$ ions can be estimated as 3:1 to reproduce the experimental spectrum.} \label{fig1}
\end{figure}
\end{center}

 A synthetic visible spectrum of W$^{26+}$ ions is shown in Fig. \ref{fig1}. The Full Width at Half Maximum (FWHM) is assumed to be 0.2nm in the Gaussian profile to reproduce the experimental spectrum.  All the four strong peaks between 330 and 390nm are from the M1 transitions between the ground multiplets of W$^{26+}$ or W$^{27+}$ ions. They were reproduced at the electron beam energies of 1100eV. The peaks that were marked as 1, 2, and 4 are the M1 transitions $^3F_4 \to {^3F_3}$ , $^3F_4 \to {^1G_4}$ , and $^3H_5 \to {^3H_4}$ from W$^{26+}$ ions, respectively. Peaks 3 is the M1 transition $^2F_{5/2} \to {^2F_{7/2}}$ of W$^{27+}$ ions. In the synthetic spectrum, the wavelength of peak 3 has been taken from the experimental observation\cite{PhysRevA.86.062501}, while the transition rate was taken from ref. \cite{0953-4075-45-3-035003}. According to the present calculation, the population ratio between W$^{26+}$ and W$^{27+}$ ions can be estimated to be 3:1 to reproduce the experimental spectrum. All the theoretical wavelength were shifted by 0.2nm to the shorter wavelength. The wavelengths and the intensities both agree well with the experimental spectrum.

The relative intensities of several strong M1 lines are given in Table \ref{tab2}. The intensities are given relative to the intensity of the transition of 389.4nm. The index of peak in the second column corresponds to the one marked in Fig. \ref{fig1}. The present theoretical relative intensities $I_{Theo}$ are given in the last column. $I_{Theo}$ are compared to the experimental data $I_{Obs}$ by Z. Fei \emph{et al.} \cite{PhysRevA.90.052517}, which are found in the fifth column of Table \ref{tab2}. All the experimental line intensities are normalized to unity for the transition at 389.4nm. The agreement between the present theory and experimental observation\cite{PhysRevA.90.052517} is satisfactory except for the line in 502.2nm from $^3 F_3\to {^3F_2}$. However, in spite of this disagreement for $^3 F_3\to {^3F_2}$ line the present theory agrees fairly well with the previous theoretical value in \cite{PhysRevA.90.052517}. This disagreement might be caused by the blending with some unknown line from other tungsten ions which is suggesting the need of more extensive experimental observation and some theoretical investigation. According to the present calculations, there are some other lines which could be observed in the future experiment, such as $^1S_0 \to {^3P_1}$, {$^3P_2 \to {^3F_3}$}, {$^1D_2 \to {^3F_3}$} located at 109.6 nm, 152.4 nm and 320.5 nm, respectively.

\begin{table}
\begin{center}
\caption{The measured and calculated relative line intensities of M1 transition in the visible region of ground state multiplets of W$^{26+}$ ions. The columns $\lambda_{Obs}$ and $I_{Obs}$ are experimental transition wavelengths and experimental intensity by Z. Fei et al\cite{PhysRevA.90.052517}. The columns $\lambda_{Theo}$ and $I_{Theo}$ are the present calculated results.}
\label{tab2}
\vspace{2mm}
\begin{tabular}{rrrrcc}
\hline
{Transition}        & {Peak}    & {$\lambda_{Obs}$(nm)}   & {$\lambda_{Theo}$(nm)}       & {$I_{Obs}$ }      & {$I_{Theo}$} \\\hline
{$^1S_0 \to {^3P_1}$}  & {}       & {}                    & {109.6}                          & {}                & {0.1}\\
{$^3P_2 \to {^3F_3}$}  & {}       & {}                    & {152.4}                          & {}                & {0.1}\\
{$^1G_4 \to {^3H_4}$}  & {}       & {263.3}               & {263.1}                          & {0.1}             & {0.2} \\
{$^3P_2 \to {^1D_2}$}  & {}       & {291.9}               & {290.9}                          & {0.1}             & {0.1} \\
{$^1D_2 \to {^3F_3}$}  & {}       & {}                    & {326.7}                          & {}                & {0.1} \\
{$^3F_4 \to {^3F_3}$}  & {1}      & {333.7}               & {333.9}                          & {0.2}             & {0.2} \\
{$^3F_4 \to {^1G_4}$}  & {2}      & {335.8}               & {335.9}                          & {0.2}             & {0.2}  \\
{$^3H_5 \to {^3H_4}$}  & {4}      & {389.4}               & {389.7}                          & {1}               & {1}   \\
{$^3H_6 \to {^3H_5}$}  & {}       & {464.6}               & {463.9}                          & {0.8}             & {0.7} \\
{$^3F_3 \to {^3F_2}$}  & {}       & {502.2}               & {501.9}                          & {1.2}             & {0.4} \\
\hline
\end{tabular}
\end{center}
\end{table}

\begin{center}
\begin{figure}
\includegraphics{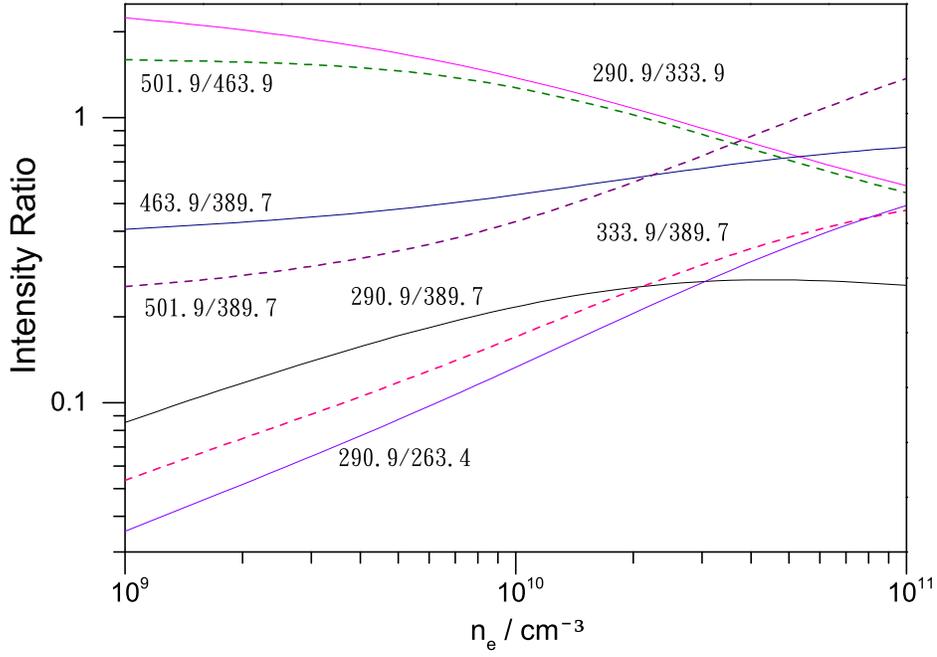}\\  
\caption{The electron density $n_e$ dependence of the emission line intensity ratios in the optical region at the electron beam energy of 1100eV. The notations "\emph{a/b}" attached to the curves represent the transition pairs which have the wavelengths \emph{a} and \emph{b}, respectively. The wavelength was taken from the column $\lambda_{Theo}$ in Table \ref{tab2}.} \label{fig2}
\end{figure}
\end{center}

The electron density $n_e$ dependence of the emission line intensity ratio from M1 transition of $4d^{10}4f^2$ of W$^{26+}$ at a mono-energetic electron beam energy of 1100eV is shown in Fig. \ref{fig2}. From this figure, it can be found that some of the selected transition pairs are sensitive to $n_e$ in the range from $10^{9}$ to $10^{11}$ cm$^{-3}$ which is a typical electron density range of the EBIT plasmas. The electron energy distribution function is assumed to be a $\delta$-function. The transition pairs were selected to be of the neighbouring in their wavelengths for the convenience of plasma diagnostics measurement. Both lines can be observed simultaneously within the range of similar instrumental efficiency, which can reduce the error of the intensity detection. It is expected that these results can be verified in EBIT experiment.

\begin{center}
\begin{figure}
\includegraphics{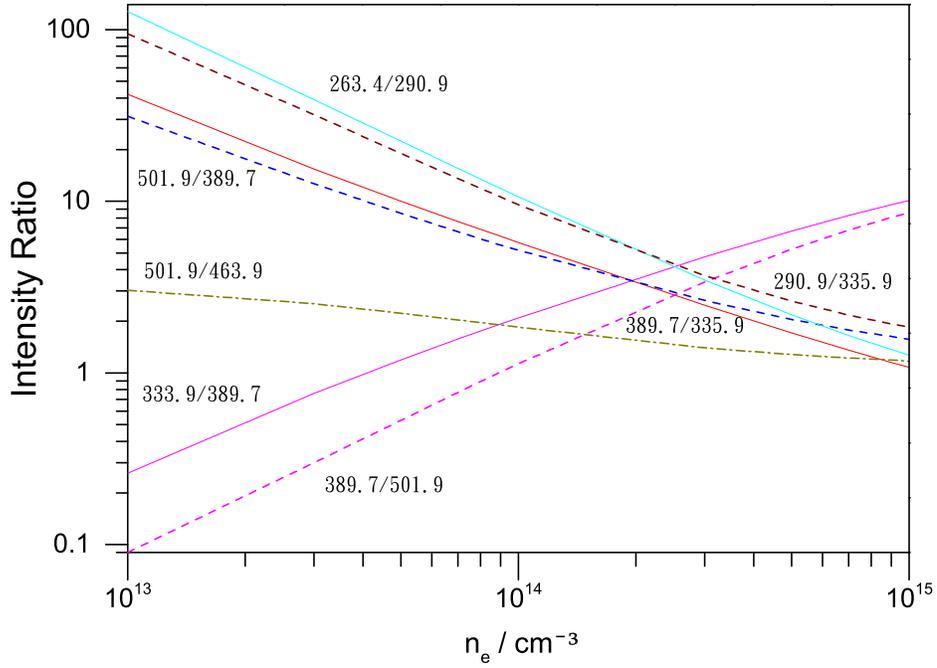}\\  
\caption{The dependence of the intensity ratio on the electron density $n_e$ for fusion plasma conditions, in which the electron energy distribution was assumed to be Maxwellian with $T_e = 1.5$ keV, $n_e$ from $10^{13}-10^{15} cm^{-3}$. The notations "\emph{a/b}" attached to the curves represent the transition pairs which have the wavelengths \emph{a} and \emph{b}, respectively. The wavelength was taken from the column $\lambda_{Theo}$ in Table \ref{tab2}.} \label{fig3}
\end{figure}
\end{center}

Furthermore, the electron density dependence of intensity ratio for fusion plasma are given in Fig. \ref{fig3}. The electron energy distribution of fusion plasma was assumed to be Maxwellian with $T_e = 1.5$ keV, $n_e$ from $10^{13}-10^{15} cm^{-3}$, which is one of the typical plasma condition for the  fusion reactor. It can be found from the figure that some intensity ratios of selected line pairs varies by about 2 orders of magnitude when the electron density varies from $10^{13}$ to $10^{15} cm^{-3}$.  The intensity ratio of these line pairs are sensitive to the electron density and expected to be observed in some fusion reactors.

\section{Conclusion}

A detailed CR model for W$^{26+}$ ions was constructed by considering the spontaneous radiative transition, electron (de)excitation processes. The presently calculated synthetic visible spectrum obtained a good agreement with the experimental one. The transition peak identification by previous work has been verified. The relative intensities for the M1 transitions have been calculated and they agree well with the experiment. Based on the present calculation, the intensity ratios between the several transition pairs were found to be sensitive to the electron density. Theoretical intensity ratios have been provided and they are expected to be examined in EBIT plasmas and possibly in the fusion plasmas. Meanwhile, the present work shows that the CR model can be efficiently used for analysis of the EBIT spectrum.

\addcontentsline{toc}{chapter}{Acknowledgment}

\section*{Acknowledgment}

This work was supported by National Nature Science Foundation of China, Grant No:11264035 and Specialized Research Fund for the Doctoral Program of Higher Education(SRFDP),Grant No: 20126203120004, International Scientific and Technological Cooperative Project of Gansu Province of China (Grant No. 1104WCGA186), JSPS-NRF-NSFC A3 Foresight Program in the field of Plasma Physics (NSFC: No.11261140328, NRF: 2012K2A2A6000443).


\section*{References}
\addcontentsline{toc}{chapter}{References}



\bibliographystyle{elsarticle-num}




\bibliography{abc}

\end{document}